\documentclass[aps,prd,preprint,superscriptaddress,showpacs,floatfix,nobibnotes]{revtex4}

\usepackage{latexsym}
\usepackage{amsmath}
\usepackage{amssymb}
\usepackage{graphicx}
\usepackage{longtable}

\usepackage{bm}
\usepackage{epsfig}
\usepackage{citesort}

\begin{document}
\title{Fluctuations and Correlations of Conserved Charges in QCD at Finite Temperature with Effective Models}

\author{Wei-jie Fu}
\email[]{wjfu@itp.ac.cn} \affiliation{Kavli Institute for
Theoretical Physics China (KITPC), Key Laboratory of Frontiers in
Theoretical Physics, Institute of Theoretical Physics, Chinese
Academy of Science, Beijing 100190, China}

\author{Yu-xin Liu}
\email[]{yxliu@pku.edu.cn} \affiliation{Department of Physics and
State Key Laboratory of Nuclear Physics and Technology, Peking
University, Beijing 100871, China} \affiliation{Center of
Theoretical Nuclear Physics, National Laboratory of Heavy Ion
Accelerator, Lanzhou 730000, China}

\author{Yue-Liang Wu}
\email[]{ylwu@itp.ac.cn} \affiliation{Kavli Institute for
Theoretical Physics China (KITPC), Key Laboratory of Frontiers in
Theoretical Physics, Institute of Theoretical Physics, Chinese
Academy of Science, Beijing 100190, China}

\date{\today}

\begin{abstract}
We study fluctuations of conserved charges including baryon number,
electric charge, and strangeness as well as the correlations among
these conserved charges in the 2+1 flavor
Polyakov--Nambu--Jona-Lasinio model at finite temperature. The
calculated results are compared with those obtained from recent
lattice calculations performed with an improved staggered fermion
action at two values of the lattice cutoff with almost physical up
and down quark masses and a physical value for the strange quark
mass. We find that our calculated results are well consistent with
those obtained in lattice calculations except for some quantitative
differences for fluctuations related with strange quarks. Our
calculations indicate that there is a pronounced cusp in the ratio
of the quartic to quadratic fluctuations of baryon number, i.e.
$\chi_{4}^{B}/\chi_{2}^{B}$, at the critical temperature during the
phase transition, which confirms that $\chi_{4}^{B}/\chi_{2}^{B}$ is
a useful probe of the deconfinement and chiral phase transition.
\end{abstract}


\maketitle

\section{Introduction}
\vspace{5pt}

QCD thermodynamics, for example the equation of state of the quark
gluon plasma (QGP), phase transition of the chiral symmetry
restoration, the deconfinement phase transition and so on, has been
a subject of intensive investigation in recent years. On the one
hand, the deconfined QGP are expected to be formed in
ultrarelativistic heavy-ion
collisions~\cite{Shuryak2004,Gyulassy2005,Shuryak2005,Arsene2005,Back2005,Adams2005,Adcox2005,Blaizot2007}
(for example the current experiments at the Relativistic Heavy Ion
Collider (RHIC) and the upcoming experiments at the Large Hadron
Collider (LHC)) and in the interior of neutron
stars~\cite{Weber2005,Alford2007,Alford2008,Fu2008b}; On the other
hand, studying the thermodynamical and hydrodynamical behaviors of
the QGP, especially the deconfinement and chiral phase transitions,
is an elementary problem in strong interaction physics.

Lattice QCD simulation is a principal approach to explore the
properties of strongly interacting matter and its deconfinement and
chiral phase transitions. In the past years, this method has
provided us with lots of information about the QCD thermodynamics
and phase transition from the confined hadronic phase to the
deconfined QGP one at finite temperature and limited chemical
potential (see, for example,
Ref.~\cite{Boyd1996,Fodor2002,Allton2002,Laermann2003,deForcrand2003,Kratochvila0456,Gavai2005,Gavai2006,Aoki2006a,Aoki2006b,Cheng2006,Ejiri2006a,Ejiri2006b,Karsch2006,Karsch2007,Cheng2009,Ejiri2009}).
In response to the lattice QCD simulations, many effective models
have been developed to describe the behavior and properties of
strongly interacting matter, give physical interpretation of the
available lattice data, and further to make predictions in the
regions of phase diagram that can not be reached by the lattice
calculations. The validity of these effective models is expected,
since what governs the critical behavior of the QCD phase transition
is the universality class of the chiral symmetry which is kept in
these effective models.

It has been known that fluctuations of conserved charges, for
example the baryon number, electric charge, and strangeness, are
particularly sensitive to the structure and behavior of the thermal
strongly interacting matter~\cite{Stephanov1998,Hatta2003,Jeon2004}.
Enhanced fluctuations are close related with the critical behavior
of the QCD thermodynamics and phase
transitions~\cite{Stephanov2004}. Furthermore, the fluctuations and
correlations of conserved charges and their high order cumulants
provide information about the degrees of freedom (confined hadrons
or deconfined QGP) of strongly interacting matter at high
temperature~\cite{Jeon2000,Koch2005}, thus they are useful probes of
the deconfinement and chiral phase
evolution~\cite{Ejiri2006a,Ejiri2006b,Karsch2006,Karsch2007,Cheng2009,Stokic2009}.
More important, the fluctuations and correlations of conserved
charges can be extracted not only theoretically from the lattice QCD
simulations and effective model calculations but also experimentally
from event-by-event
fluctuations~\cite{Jeon2004,Koch2005,Abelev2009}.

In this work, we will study the fluctuations and correlations of
conserved charges and their high order cumulants in the 2+1 flavor
Polyakov--Nambu--Jona-Lasinio (PNJL) model~\cite{Fu2008}. Compared
with the conventional Nambu--Jona-Lasinio model, the PNJL model not
only has the chiral symmetry as well as the dynamical breaking
mechanism of this symmetry, but also include the effect of color
confinement through the Polyakov
loop~\cite{Meisinger9602,Pisarski2000,Fukushima2004,Ratti2006a,Ratti2006b,Ciminale2008,Zhang2006,Fu2009,Ghosh2006}.
Our calculated results of the fluctuations and correlations of
conserved charges will be compared with those obtained from very
recent lattice calculations performed with an improved staggered
fermion action at two values of the lattice cutoff with almost
physical up and down quark masses and a physical value for the
strange quark mass~\cite{Cheng2009}. On the one hand, our
calculations provide insight about what can not be concluded from
lattice calculations. For example, whether there is a cusp in the
ratio of the quartic to quadratic fluctuations of baryon number
$\chi_{4}^{B}/\chi_{2}^{B}$, which is a valuable probe of the
deconfinement and chiral dynamics, at the critical temperature can
not yet be answered by lattice simulations, because of the larger
errors in the calculations, but our calculations indeed find a
pronounced cusp in this ratio $\chi_{4}^{B}/\chi_{2}^{B}$ at the
critical temperature. Furthermore, our calculations give physical
interpretation of the lattice data. On the other hand, comparing our
calculated results with those obtained in lattice calculations, it
allows us to test the validity of the 2+1 flavor PNJL model, and
then provide the right direction for improving this effective model.

The paper is organized as follows. In Sec. II we simply review the
formalism of the 2+1 flavor PNJL model. In Sec. III we introduce the
fluctuations and correlations of conserved charges. In Sec. IV we
give our calculated results of fluctuations of light quarks (up and
down quarks) and strange quarks. In Sec. V we give our calculated
results of fluctuations and correlations of conserved charges, and
compare them with those obtained in lattice calculations. In Sec. VI
we present our summary and conclusions.

\section{2+1 flavor PNJL model}

In this work, we employ the 2+1 flavor Polyakov-loop improved NJL
model which has been discussed in details in our previous work
~\cite{Fu2008}, and the Lagrangian density for the 2+1 flavor PNJL
model is given as
\begin{eqnarray}
\mathcal{L}_{PNJL}&=&\bar{\psi}\left(i\gamma_{\mu}D^{\mu}+\gamma_{0}
 \hat{\mu}-\hat{m}_{0}\right)\psi
 +G\sum_{a=0}^{8}\left[\left(\bar{\psi}\tau_{a}\psi\right)^{2}
 +\left(\bar{\psi}i\gamma_{5}\tau_{a}\psi\right)^{2}\right]\notag   \nonumber \\
&&-K\left[\textrm{det}_{f}\left(\bar{\psi}\left(1+\gamma_{5}\right)\psi\right)
 +\textrm{det}_{f}\left(\bar{\psi}\left(1-\gamma_{5}\right)\psi\right)\right]
 -\mathcal{U}\left(\Phi,\Phi^{*} \, ,T\right),\label{lagragian}
\end{eqnarray}
where $\psi=(\psi_{u},\psi_{d},\psi_{s})^{T}$ is the three-flavor
quark field,
\begin{equation}
D^{\mu}=\partial^{\mu}-iA^{\mu}\quad\textrm{with}\quad
A^{\mu}=\delta^{\mu}_{0}A^{0}\quad\textrm{,}\quad
A^{0}=g\mathcal{A}^{0}_{a}\frac{\lambda_{a}}{2}=-iA_4.
\end{equation}
$\lambda_{a}$ are the Gell-Mann matrices in color space and the
gauge coupling $g$ is combined with the SU(3) gauge field
$\mathcal{A}^{\mu}_{a}(x)$ to define $A^{\mu}(x)$ for convenience.
$\hat{m}_{0}=\textrm{diag}(m_{0}^{u},m_{0}^{d},m_{0}^{s})$ is the
three-flavor current quark mass matrix. Throughout this work, we
take $m_{0}^{u}=m_{0}^{d}\equiv m_{0}^{l}$, while keep $m_{0}^{s}$
being larger than $m_{0}^{l}$, which breaks the $SU(3)_f$ symmetry.
The quark chemical potentials are contained in the matrix
$\hat{\mu}=\textrm{diag}(\mu_{u}, \mu_{d}, \mu_{s})$, and they can
also be expressed in terms of chemical potentials for baryon number
$(\mu_{B})$, electric charge $(\mu_{Q})$, and strangeness
$(\mu_{S})$:
\begin{equation}
\mu_{u}=\frac{1}{3}\mu_{B}+\frac{2}{3}\mu_{Q},\quad
\mu_{d}=\frac{1}{3}\mu_{B}-\frac{1}{3}\mu_{Q},\quad\textrm{and}\quad
\mu_{s}=\frac{1}{3}\mu_{B}-\frac{1}{3}\mu_{Q}-\mu_{S}.
\end{equation}

In the above PNJL Lagrangian,
$\mathcal{U}\left(\Phi,\Phi^{*},T\right)$ is the Polyakov-loop
effective potential, which is expressed in terms of the traced
Polyakov-loop $\Phi=(\mathrm{Tr}_{c}L)/N_{c}$ and its conjugate
$\Phi^{*}=(\mathrm{Tr}_{c}L^{\dag})/N_{c}$ with the Polyakov-loop
$L$ being a matrix in color space given explicitly by
\begin{equation}
L\left(\vec{x}\right)=\mathcal{P}\exp\left[i\int_{0}^{\beta}d\tau\,
A_{4}\left(\vec{x},\tau\right)\right] =\exp\left[i \beta A_{4}
\right]\, ,
\end{equation}
with  $\beta=1/T$ being the inverse of temperature and
$A_{4}=iA^{0}$.

In our work, we use the Polyakov-loop effective potential which is a
polynomial in $\Phi$ and $\Phi^{*}$~\cite{Ratti2006a}, given by

\begin{equation}
\frac{\mathcal{U}\left(\Phi,\Phi^{*},T\right)}{T^{4}} =
-\frac{b_{2}(T)}{2}\Phi^{*}\Phi -\frac{b_{3}}{6}
(\Phi^{3}+{\Phi^{*}}^{3})+\frac{b_{4}}{4}(\Phi^{*}\Phi)^{2} \, ,
\end{equation}
with
\begin{equation}
b_{2}(T)=a_{0}+a_{1}\left(\frac{T_{0}}{T}\right)+a_{2}
{\left(\frac{T_{0}}{T}\right)}^{2}
+a_{3}{\left(\frac{T_{0}}{T}\right)}^{3}.
\end{equation}
Parameters in the effective potential are fitted to reproduce the
thermodynamical behavior of the pure gauge QCD obtained from the
lattice simulations, and their values are given in
Table~\ref{pol_para}. The parameter $T_{0}$ is the critical
temperature for the deconfinement phase transition to take place in
the pure-gauge QCD and $T_{0}$ is chosen to be $270\,\mathrm{MeV}$
according to the lattice calculations.

\begin{table}[!htb]
\begin{center}
\caption{Parameters for the Polyakov-loop effective potential
$\mathcal{U}$} \label{pol_para}
\begin{tabular}{cccccc}
\hline \hline \vspace{0.1cm}
$a_{0}$\qquad\qquad&$a_{1}$\qquad\qquad&$a_{2}$\qquad\qquad&$a_{3}$\qquad\qquad&
$b_{3}$\qquad\qquad& $b_{4}$\\
\hline 6.75 \qquad\qquad& $-1.95$ \qquad\qquad& 2.625\qquad\qquad& $-7.44$\qquad\qquad& 0.75\qquad\qquad& 7.5\\
\hline
\end{tabular}
\end{center}
\end{table}

In the mean field approximation, the thermodynamical potential
density for the 2+1 flavor quark system is given by
\begin{eqnarray}
\Omega&=&-2N_{c}\sum_{f=u,d,s}\int\frac{d^{3}p}{\left(2\pi\right)^{3}}\bigg\{E_{p}^{f}\theta(\Lambda^{2}-p^{2})\nonumber \\
&&+ \frac{T}{3}\ln\Big[1+3\Phi^{*}e^{-(E_{p}^{f}-\mu_{f})/T}+3\Phi
e^{-2(E_{p}^{f}-\mu_{f})/T}+e^{-3(E_{p}^{f}-\mu_{f})/T}\Big]
\nonumber \\
&&+ \frac{T}{3}\ln\Big[1+3\Phi e^{-(E_{p}^{f}+\mu_{f})/T}+3\Phi^{*}
e^{-2(E_{p}^{f}+\mu_{f})/T}+e^{-3(E_{p}^{f}+\mu_{f})/T}\Big]\bigg\}\nonumber \\
&&+2G({\phi_{u}}^{2}
+{\phi_{d}}^{2}+{\phi_{s}}^{2})-4K\phi_{u}\,\phi_{d}\,\phi_{s}+\mathcal{U}(\Phi,\Phi^{*},T).\label{thermopotential}
\end{eqnarray}

where $\phi_{i}\,(i=u,d,s)$ is the chiral condensate of quarks with
flavor $i$, and we have the energy-momentum dispersion relation
$E_{p}^{i}=\sqrt{p^{2}+M_{i}^{2}}$ for its corresponding
quasiparticle, with the constituent mass for the quark of flavor $i$
being
\begin{equation}
M_{i}=m_{0}^{i}-4G\phi_{i}+2K\phi_{j}\,\phi_{k}.\label{constituentmass}
\end{equation}

Minimizing the thermodynamical potential in
Eq.~\eqref{thermopotential} with respective to $\phi_{u}$,
$\phi_{d}$, $\phi_{s}$, $\Phi$, and $\Phi^{*}$, we obtain a set of
equations of motion
\begin{equation}
\frac{\partial\Omega}{\partial\phi_{u}}=0, \quad
\frac{\partial\Omega}{\partial\phi_{d}}=0, \quad
\frac{\partial\Omega}{\partial\phi_{s}}=0, \quad
\frac{\partial\Omega}{\partial\Phi}=0, \quad
\frac{\partial\Omega}{\partial\Phi^{*}}=0.\label{motion_equations}
\end{equation}
Then, this set of equations can be solved as functions of
temperature $T$ and three flavor quark chemical potentials
$\mu_{u}$, $\mu_{d}$, and $\mu_{s}$ or chemical potentials of
conserved charges $\mu_{B}$, $\mu_{Q}$, and $\mu_{S}$.

\section{Fluctuations and Correlations}

Following the procedure in lattice calculations~\cite{Cheng2009}, we
focus on the derivatives of the pressure ($P=-\Omega$) of the
thermodynamical system with respective to the chemical potentials
corresponding to the conserved charge: baryon number, electric
charge, and strangeness, i.e.
\begin{equation}
\chi_{ijk}^{BQS}=\frac{\partial^{i+j+k}(P/T^{4})}
{\partial(\mu_{B}/T)^{i}\partial(\mu_{Q}/T)^{j}\partial(\mu_{S}/T)^{k}}\bigg|
_{\mu_{B,Q,S}=0},\label{susceptibility}
\end{equation}
which are evaluated at $\mu_{B,Q,S}=0$. The $\chi$'s in
Eq.~\eqref{susceptibility} are in fact the generalized
susceptibilities and they are related with the moments of charge
fluctuations, i.e. $\delta N_{X}\equiv N_{X}-\langle N_{X}\rangle$
($X=B,Q,S$) and the correlations among conserved charges. Taking the
quadratic, quartic and the 6th order charge fluctuations for
example, we have
\begin{eqnarray}
\chi_{2}^{X}&=&\frac{1}{VT^{3}}\langle {\delta N_{X}}^{2} \rangle,  \\
\chi_{4}^{X}&=&\frac{1}{VT^{3}}\Big(\langle {\delta
N_{X}}^{4}\rangle-3{\langle {\delta N_{X}}^{2} \rangle}^{2}\Big), \\
\chi_{6}^{X}&=&\frac{1}{VT^{3}}\Big(\langle {\delta
N_{X}}^{6}\rangle-15\langle {\delta N_{X}}^{4}\rangle\langle {\delta
N_{X}}^{2}\rangle-10{\langle {\delta
N_{X}}^{3}\rangle}^{2}+30{\langle {\delta
N_{X}}^{2}\rangle}^{3}\Big).
\end{eqnarray}
And the correlations among two conserved charges are
\begin{equation}
\chi_{11}^{XY}=\frac{1}{VT^{3}}\langle N_{X}N_{Y}\rangle.
\end{equation}
It should be noted that the fluctuations and correlations of
conserved charges, i.e. the generalized susceptibilities in
Eq.~\eqref{susceptibility} are nonvanishing only when $i+j+k$ is
even at vanishing chemical potentials $\mu_{B,Q,S}=0$.

In this work we use the method of Taylor expansion to compute the
fluctuations and correlations of conserved charges in the PNJL
model. Before our numerical calculations, we should determine the
five parameters in the quark sector of the model. Values of the five
parameters used usually in the literatures are obtained in
Ref.~\cite{Rehberg1996}, which are $m_{0}^{l}=5.5\;\mathrm{MeV}$,
$m_{0}^{s}=140.7\;\mathrm{MeV}$, $G\Lambda^{2}=1.835$,
$K\Lambda^{5}=12.36$ and $\Lambda=602.3\;\mathrm{MeV}$. They are
fixed by fitting $m_{\pi}=135.0\;\mathrm{MeV}$,
$m_{K}=497.7\;\mathrm{MeV}$, $m_{\eta^{\prime}}=957.8\;\mathrm{MeV}$
and $f_{\pi}=92.4\;\mathrm{MeV}$. For convenience to compare our
calculations with the recent lattice simulations in which the
strange quark mass has been tuned close to its physical value and
the light quark masses have been chosen to be one tenth of the
strange quark mass~\cite{Cheng2009}, we employ the values of the
five parameters given above except with
$m_{0}^{l}=14.0\;\mathrm{MeV}$, which is consistent with the lattice
calculations.

\section{Fluctuations of Light Quarks and Strange Quarks}

\begin{figure}[!htb]
\includegraphics[scale=1.2]{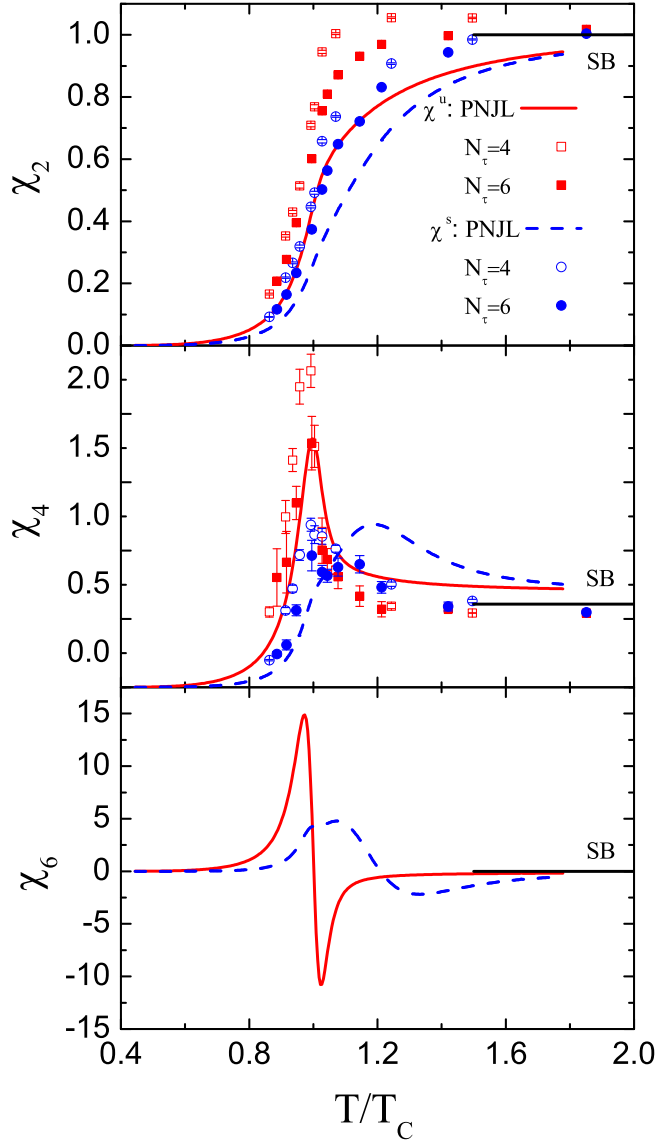}
\caption{(color online). Quadratic (top), quartic (middle) and the
6th order (bottom) fluctuations of light quarks (here we denote them
with $\chi^{u}$) and strange quarks as functions of the temperature
(in unit of the critical temperature $T_{C}$) calculated in the PNJL
model with $m_{0}^{l}=14.0\;\mathrm{MeV}$ and
$m_{0}^{s}=140.7\;\mathrm{MeV}$. We also show the results of the
quadratic and quartic fluctuations obtained from calculations on
$16^{3}\times4$ and $24^{3}\times6$ lattices in
Ref.~\cite{Cheng2009}. Here SB denotes the corresponding limit value
of Stefan-Boltzmann ideal gas.}\label{f1}
\end{figure}

In order to understand the behavior of the fluctuations of conserved
charges better, we will study the fluctuations of light quarks and
strange quarks in this section. In Fig.~\ref{f1} we show the
quadratic, quartic and the 6th order fluctuations of light and
strange quarks versus temperature. We find that the pseudo-critical
temperature for the chiral restoration phase transition is
$225\;\mathrm{MeV}$ and that for the deconfinement phase transition
is $221\;\mathrm{MeV}$ at vanishing chemical potential in the PNJL
model with $m_{0}^{l}=14.0\;\mathrm{MeV}$,
$m_{0}^{s}=140.7\;\mathrm{MeV}$ and other parameters given above.
More detailed discussions about the different phase transitions can
be found in Ref~\cite{Fu2008}. Since these two critical temperatures
for the two different kinds of phase transitions are almost same, we
just need one value of temperature to locate the two phase
transitions and here we choose $T_{C}=225\;\mathrm{MeV}$ which is
used as unit in Fig.~\ref{f1}. Furthermore, we also present the
results of the quadratic and quartic fluctuations obtained from
calculations on $16^{3}\times4$ and $24^{3}\times6$ lattices in
Ref~\cite{Cheng2009} in Fig.~\ref{f1}. From the top panel of
Fig.~\ref{f1} ones can find that the quadratic fluctuations
calculated in the PNJL model increase monotonously with the increase
of the temperature, and the fluctuations of the heavier strange
quarks are suppressed relative to those of the light quarks. These
features are consistent with the lattice calculations (also the
lattice simulations in Ref.~\cite{Gavai2006}). Comparing our results
with those from lattice calculations, we find that the quadratic
fluctuations calculated in the PNJL model are relatively smaller
than those calculated in the lattice simulations. And we also find
that the quadratic fluctuations calculated in the lattice
simulations rapidly approach the Stefan-Boltzmann ideal massless gas
value, i.e. $\chi_{2}^{SB}=1$ when the temperature is above the
critical temperature, while the quadratic fluctuations calculated in
the PNJL model approach the Stefan-Boltzmann value more slowly,
which is because the current quark mass effect is obvious at high
temperature in the PNJL model.

\begin{figure}[!htb]
\includegraphics[scale=0.5]{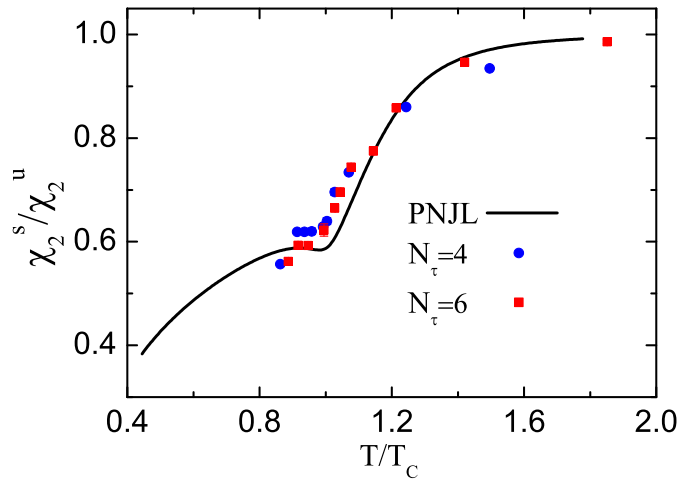}
\caption{(color online). Ratio of $s$ and $u$ quark quadratic
fluctuations as function of the temperature calculated in the PNJL
model with $m_{0}^{l}=14.0\;\mathrm{MeV}$ and
$m_{0}^{s}=140.7\;\mathrm{MeV}$ and in simulations on
$16^{3}\times4$ and $24^{3}\times6$ lattices in
Ref.~\cite{Cheng2009}.}\label{f5}
\end{figure}

In Fig.~\ref{f5} we show the ratio of strange and light quark
quadratic fluctuations as function of the temperature. Our
calculations clearly indicate that the quadratic fluctuations of
strange quarks are smaller than those of light quarks at low
temperature, which are consistent with the lattice calculations.
Furthermore, we find a plateau at the critical temperature in the
curve of $\chi^{s}_{2}/\chi^{u}_{2}$ versus temperature in our
calculations, as the solid line shows in Fig.~\ref{f5}. This plateau
can also be found in the lattice data.

\begin{figure}[!htb]
\includegraphics[scale=0.5]{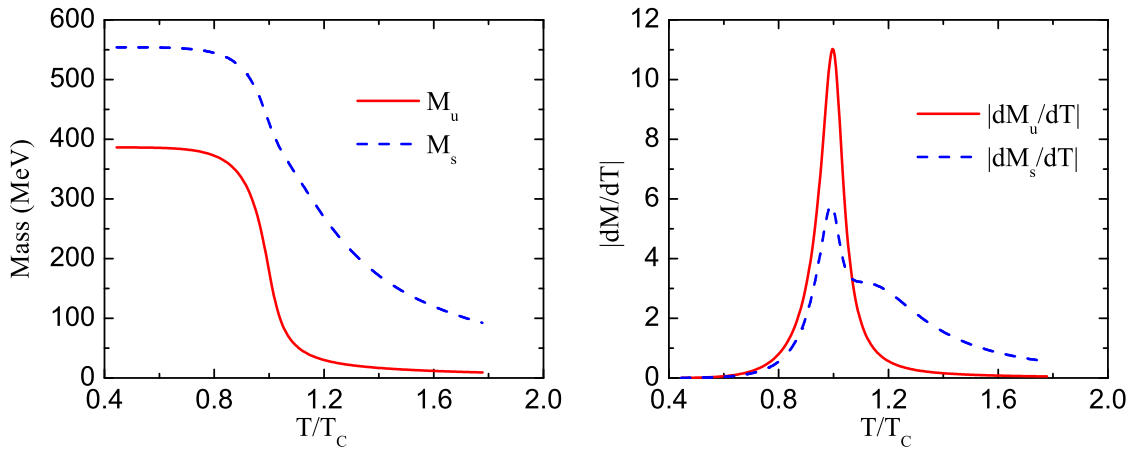}
\caption{(color online). Left panel: constituent masses of light
quarks and strange quarks as functions of the temperature at
vanishing chemical potentials in the PNJL model with
$m_{0}^{l}=14.0\;\mathrm{MeV}$ and $m_{0}^{s}=140.7\;\mathrm{MeV}$.
Right panel: corresponding derivatives of constituent masses of
light quarks and strange quarks with respective to the temperature,
and here we use the absolute values of the derivatives.}\label{f2}
\end{figure}

As for the quartic fluctuations, we find that the fluctuations of
light quarks calculated in the PNJL model is consistent with those
obtained from the lattice calculations, especially during the phase
transition where a sharp peak appears at $T_{C}$. Furthermore,
lattice calculations show that there is a relatively lower peak at
$T_{C}$ as well for the quartic fluctuations of strange quarks,
while our calculations indicate that the peak corresponding to the
strange quarks is shifted to about $1.2T_{C}$ and becomes broader
and lower. To explore the reason for the difference between our and
lattice calculations, we show the constituent masses of light quarks
and strange quarks, and also their derivatives with respective to
temperature as functions of the temperature in Fig.~\ref{f2}. One
can find that the constituent mass of light quarks rapidly decreases
to their current quark mass at $T_{C}$, which corresponds to a sharp
peak of the curve $|dM_{u}/dT|$ at the critical temperature as the
right panel of Fig.~\ref{f2} shows. However, for the strange quarks,
the constituent mass still decreases slowly beyond the critical
temperature, which results in a plateau of the curve $|dM_{s}/dT|$
at about $1.2T_{C}$. That is the reason why the quartic fluctuations
of strange quarks present a broad peak at $1.2T_{C}$ in the PNJL
model. In the bottom panel of Fig.~\ref{f1} we show the 6th order
fluctuations of light quarks and strange quarks. One can find that
the 6th order fluctuations of light quarks oscillate violently at
$T_{C}$, while those of strange quarks oscillate mildly at
relatively larger temperature.

\begin{figure}[!htb]
\includegraphics[scale=0.5]{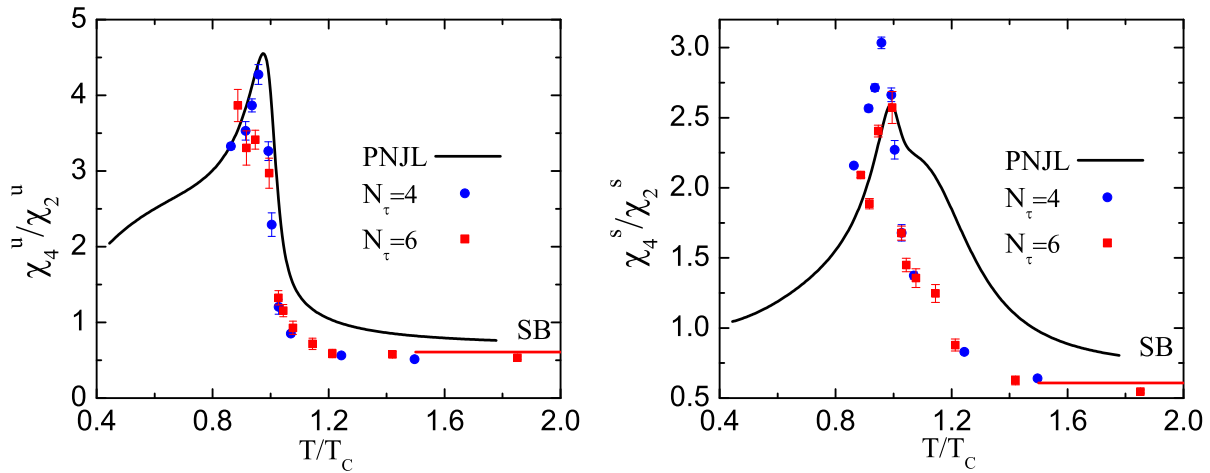}
\caption{(color online). Ratio of the quartic to quadratic
fluctuations for light quarks (left panel) and strange quarks (right
panel) as functions of the temperature calculated in the PNJL model
with $m_{0}^{l}=14.0\;\mathrm{MeV}$ and
$m_{0}^{s}=140.7\;\mathrm{MeV}$, which are also compared with those
obtained from calculations on $16^{3}\times4$ and $24^{3}\times6$
lattices in Ref.~\cite{Cheng2009}.}\label{f3}
\end{figure}

In the following we pay more attentions to the ratio of the quartic
to quadratic fluctuations of the quark number, since this ratio is
believed to be a valuable probe of the deconfinement and chiral
phase
transitions~\cite{Ejiri2006a,Karsch2006,Ejiri2006b,Stokic2009}.
Fig.~\ref{f3} shows this ratio for the light quarks and strange
quarks calculated in our PNJL model, and we also show the
corresponding results obtained in the lattice simulations for
comparisons. Our calculations indicate that, for both light quarks
and strange quarks, the ratio of the quartic to quadratic
fluctuations of the quark number increases with the temperature
below the critical temperature, while decreases when the temperature
is above the critical one. So there is a peak in the curve of the
ratio versus temperature at the critical temperature for both light
quarks and strange quarks. These features are consistent with those
obtained in the lattice calculations. Comparing the two panels of
Fig.~\ref{f3} we find that the consistency between our calculations
and the lattice simulations for the light quarks is better than that
for the strange quarks, which can be expected, since for the quartic
fluctuations of strange quarks the difference between the PNJL model
and the lattice simulations is quite larger than that for the light
quarks, as shown in the middle panel of Fig.~\ref{f1}.

Next, we discuss the ratio of the quartic to quadratic fluctuations
when the temperature is away from the phase transition temperature.
As the temperature is high, especially when the temperature is quite
above the critical temperature, the system can be approximated as
noninteracting massless gases. That is the Stefan-Boltzmann limit,
and in this limit the pressure of the quarks and antiquarks can be
easily obtained as
\begin{equation}
\frac{P}{T^{4}}=\sum_{f=u,d,s}N_{c}\bigg[\frac{7\pi^{2}}{180}+
\frac{1}{6}\Big(\frac{\mu_{f}}{T}\Big)^{2}+\frac{1}{12\pi^{2}}\Big(\frac{\mu_{f}}{T}\Big)^{4}\bigg].\label{pres_highT}
\end{equation}
So in the Stefan-Boltzmann limit, $\chi_{4}/\chi_{2}=6/\pi^{2}$.

\begin{figure}[!htb]
\includegraphics[scale=0.5]{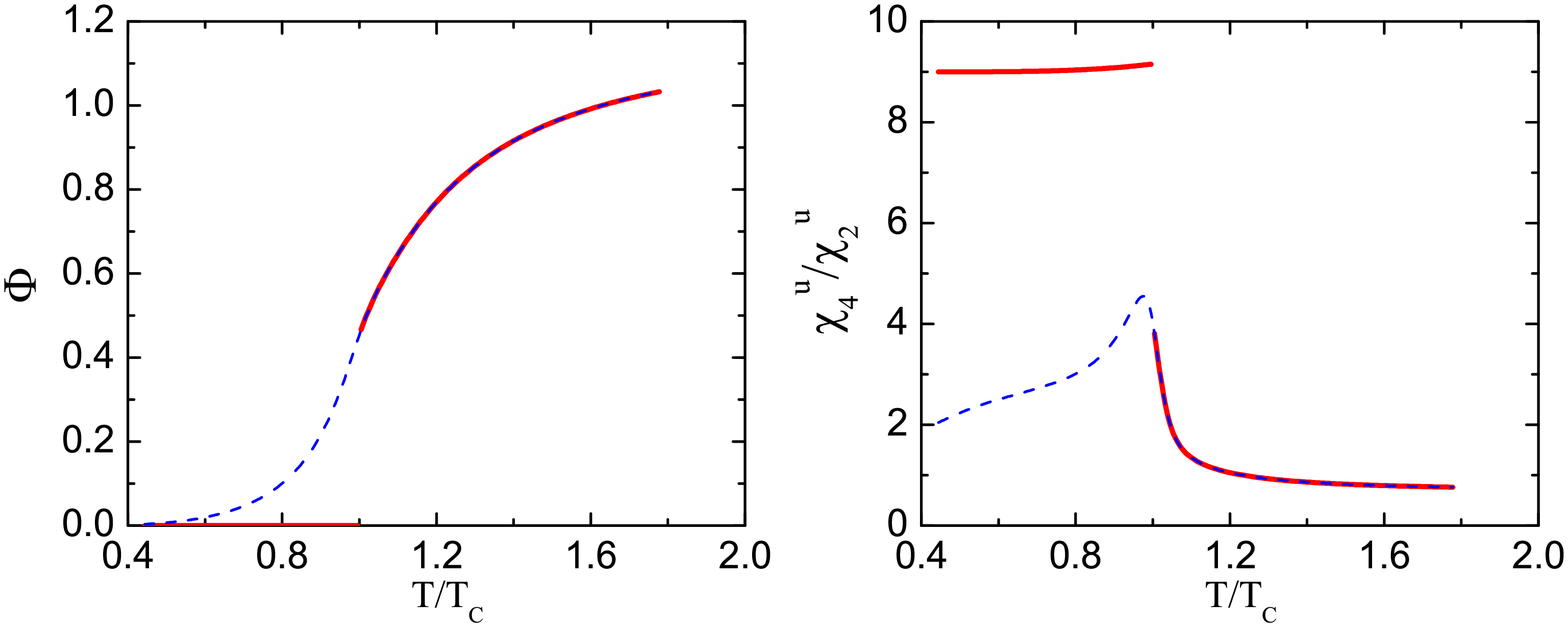}
\caption{(color online). Left panel: Polyakov-loop versus
temperature in two cases at vanishing chemical potentials in the
PNJL model with $m_{0}^{l}=14.0\;\mathrm{MeV}$ and
$m_{0}^{s}=140.7\;\mathrm{MeV}$ ($\Phi^{*}=\Phi$ at vanishing
chemical potentials). The blue dashed line corresponds to our full
calculations, and the red solid line to the calculations in which
$\Phi$ and $\Phi^{*}$ are set to be zero artificially when the
temperature is below the critical temperature. Right panel: ratio of
the quartic to quadratic fluctuations for light quarks corresponding
to the two cases. Therefore, the blue dashed line represents the
same results which have been shown in the left panel of
Fig.~\ref{f3}.}\label{f4}
\end{figure}

On the contrary, when the temperature is below the critical
temperature, the constituent masses of quarks are much larger than
their current masses as the left panel of Fig.~\ref{f2} shows.
Furthermore, quarks tend to be combined to form colorless hadrons,
which is the color confinement. In the PNJL model, we can not
include the degrees of freedom of hadrons, but this effective model
embodies the confinement effect by suppressing the excitations of
the one and two quark states and only permitting those of three
quark states~\cite{Fu2008}. In details, when the temperature is much
lower than the critical temperature, the Polyakov-loop and its
conjugate $\Phi,\Phi^{*}\rightarrow 0$, and then most contributions
to the pressure of the thermodynamical system come from the three
quark states. Since the constituent masses of quarks are much larger
than the temperature, we can use the Boltzmann
approximation~\cite{Stokic2009}, and the pressure is
\begin{equation}
\frac{P}{T^{4}}\simeq\sum_{f=u,d,s}N_{c}\frac{2}{81\pi^{2}}\Big(\frac{3m_{f}}{T}\Big)^{2}K_{2}\Big(\frac{3m_{f}}{T}\Big)
\cosh\Big(\frac{3\mu_{f}}{T}\Big).\label{pres_lowT}
\end{equation}
Where $K_{2}$ is a Bessel function. Then we can easily obtain
$\chi_{4}/\chi_{2}=9$. It should be emphasized that, since the
temperature driven deconfinement transition is not a strict phase
transition but a continuous crossover~\cite{Aoki2006a,Aoki2006b},
the Polyakov-loop $\Phi$ (or $\Phi^{*}$) is not vanishing when the
temperature is below the critical one as the blue dashed line shows
in the left panel of Fig.~\ref{f4}. Assuming $\Phi=\Phi^{*}=1$ even
when $T<T_{C}$, we obtain
\begin{equation}
\frac{P}{T^{4}}\simeq\sum_{f=u,d,s}N_{c}\frac{2}{\pi^{2}}\Big(\frac{m_{f}}{T}\Big)^{2}K_{2}\Big(\frac{m_{f}}{T}\Big)
\cosh\Big(\frac{\mu_{f}}{T}\Big).\label{pres_lowT1}
\end{equation}
Then we have $\chi_{4}/\chi_{2}=1$ in this situation. Since $\Phi$
(or $\Phi^{*}$) is between 0 and 1 at low temperature as we have
mentioned above, we expect $\chi_{4}/\chi_{2}$ to be between 1 and 9
when $T<T_{C}$. This expectation is verified in our numerical
calculations as shown in Fig.~\ref{f3}. For comparison, we set
$\Phi$ and $\Phi^{*}$ to be zero artificially when the temperature
is below the critical temperature as the red solid line shows in the
left panel of Fig.~\ref{f4}, then we calculate the corresponding
$\chi^{u}_{4}/\chi^{u}_{2}$. The results are shown in the right
panel of Fig.~\ref{f4}, which clearly indicates that
$\chi^{u}_{4}/\chi^{u}_{2}$ approaches 9 when $T<T_{C}$. One can
easily find that the ratio of the quartic to quadratic fluctuations
calculated in the case in which the Polyakov-loop is set to be zero
artificially is not consistent with the lattice calculations.

\begin{figure}[!htb]
\includegraphics[scale=0.5]{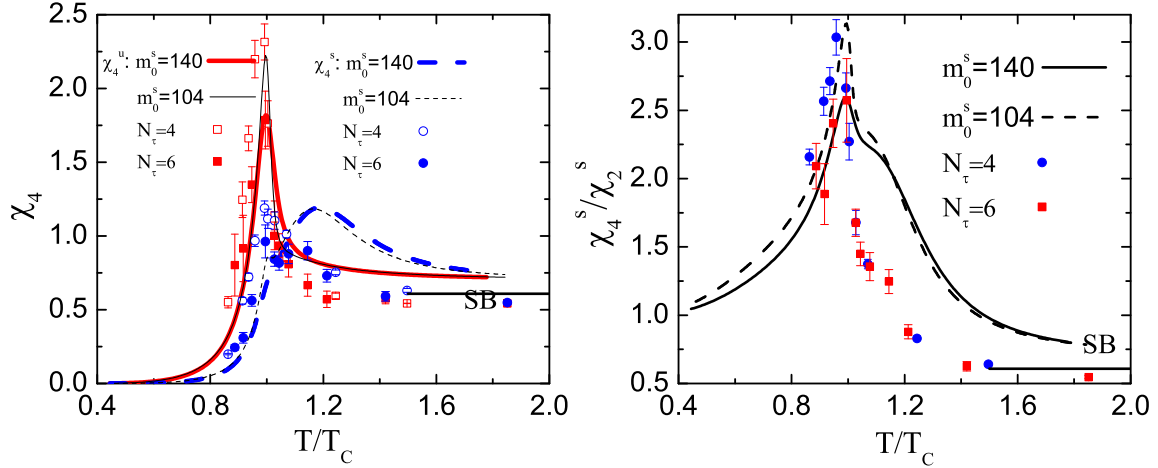}
\caption{(color online). Left panel: quartic fluctuations of light
quarks and strange quarks as functions of the temperature in the
PNJL model with different quark current masses, one with
$m_{0}^{l}=14.0\;\mathrm{MeV}$ and $m_{0}^{s}=140.7\;\mathrm{MeV}$
(same as our previous calculations) and the other with
$m_{0}^{l}=3.79\;\mathrm{MeV}$ and $m_{0}^{s}=104\;\mathrm{MeV}$. In
the same way, the corresponding results obtained in QCD simulations
on $16^{3}\times4$ and $24^{3}\times6$ lattices in
Ref.~\cite{Cheng2009} are also presented. Right panel: ratio of the
quartic to quadratic fluctuations for strange quarks as functions of
the temperature in the PNJL model with the two sets of current quark
masses.}\label{f6}
\end{figure}

So far, we have found that the calculations in the PNJL model for
the light quark fluctuations are consistent with those in the
lattice simulations, while for the strange quark fluctuations, since
the current mass of the strange quarks is large, there are
relatively larger differences between the results in the PNJL model
and those in lattice calculations. Considering the value of the
strange quark current mass used in the PNJL model
($m_{0}^{s}=140.7\;\mathrm{MeV}$) is relatively larger than the
middle value $104\;\mathrm{MeV}$ given by the PDG2008
($m_{0}^{s}=104^{+26}_{-34}\;\mathrm{MeV}$~\cite{Amsler2008}), we
set $m_{0}^{s}=104\;\mathrm{MeV}$ and $m_{0}^{l}=3.79\;\mathrm{MeV}$
in the PNJL model and leave other parameters unchanged (here
$m_{0}^{l}=3.79\;\mathrm{MeV}$ is the middle value of the average
mass of up and down quarks given by the PDG2008
($\overline{m}=(m_{u}+m_{d})/2=3.79^{+1.21}_{-1.29}\;\mathrm{MeV}$~\cite{Amsler2008})).
Then we repeat the calculations above. Fig.~\ref{f6} shows the
quartic fluctuations of light quarks and strange quarks, and the
ratio of the quartic to quadratic fluctuations for strange quarks
with the two sets of current quark masses. We find that the height
of the peak in the quartic fluctuations of light quarks at the
critical temperature is enhanced due to the reduction of light quark
current mass, and with the decrease of the strange quark mass, the
broad peak of the curve $\chi_{4}^{s}$ move a little to low
temperature. However, there are also differences between the PNJL
model with $m_{0}^{s}=104\;\mathrm{MeV}$ and the lattice simulations
for the $\chi_{4}^{s}$ and $\chi_{4}^{s}/\chi_{2}^{s}$.

\section{Fluctuations and Correlations of Conserved Charges}

\begin{figure}[!htb]
\includegraphics[scale=1.2]{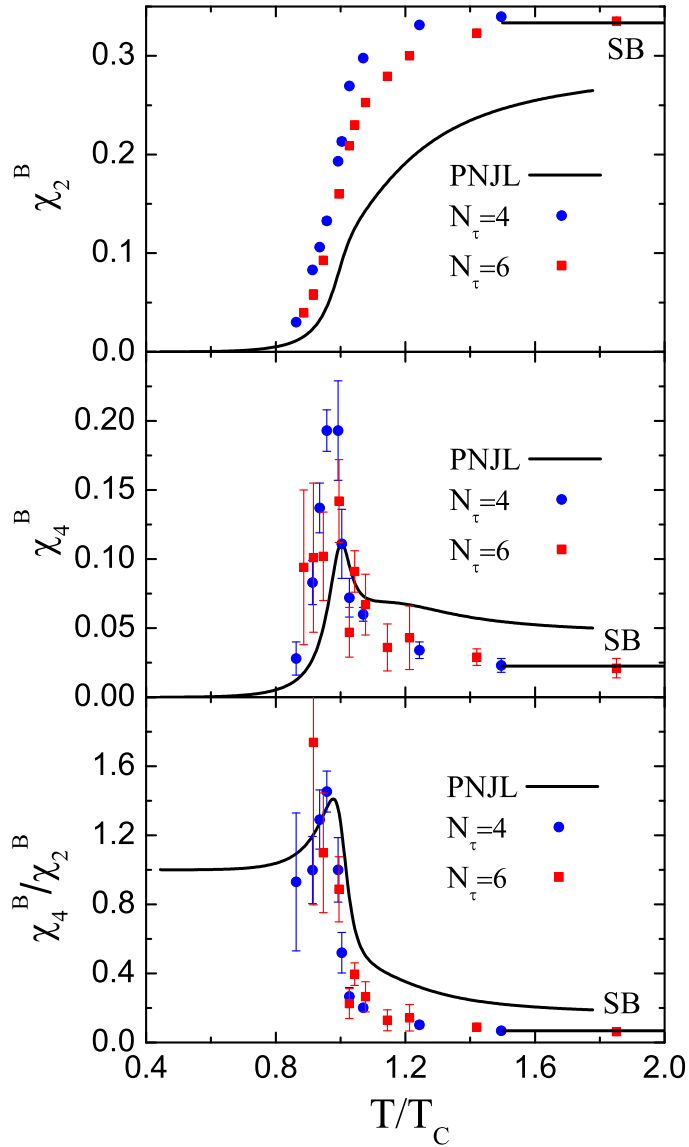}
\caption{(color online). Quadratic (top), quartic (middle)
fluctuations of baryon number, and their ratio (bottom) as functions
of the temperature calculated in the PNJL model with
$m_{0}^{l}=14.0\;\mathrm{MeV}$ and $m_{0}^{s}=140.7\;\mathrm{MeV}$.
The corresponding results obtained from calculations on
$16^{3}\times4$ and $24^{3}\times6$ lattices in
Ref.~\cite{Cheng2009} are also presented.}\label{f7}
\end{figure}

We have discussed the fluctuations of light quarks and strange
quarks in details above. In the following, we will focus on the
fluctuations of baryon number, electric charge, and strangeness.

Fig.~\ref{f7} shows the quadratic and quartic fluctuations of the
baryon number, and their ratio versus temperature calculated in the
PNJL model and in the lattice simulations. Same as the fluctuations
of light quarks and strange quarks, We find that the quadratic
fluctuations of baryon number increase monotonically with the
temperature, while the quartic fluctuations develop a cusp at the
critical temperature and decrease with increase of the temperature
when $T>T_{C}$. As a matter of fact, the singular behavior of the
baryon number fluctuations is expected to be controlled by the
universal $O(4)$ symmetry group at vanishing chemical potential and
vanishing light quark mass~\cite{Hatta2003}. And the baryon number
fluctuations are expected to scale like~\cite{Cheng2009},
\begin{equation}
\chi_{2n}^{B}\sim
\Big|\frac{T-T_{C}}{T_{C}}\Big|^{2-n-\alpha}\label{chiBT}
\end{equation}
with $\alpha\simeq-0.25$. As Eq.~\eqref{chiBT} shows, with the
increase of the order, the singular properties of the baryon number
fluctuations become more and more prominent, which are confirmed
both in the calculations of the PNJL model and in the lattice
simulations. Comparing the results obtained in the PNJL model and
those in lattice calculations, we find that our results are
consistent with the lattice results qualitatively, but the
differences in quantities between the two results still exist,
especially at high temperature. The reason for these differences has
been mentioned above, i.e. when the temperature is high, the current
quark mass effect becomes more and more important.

The bottom panel of Fig.~\ref{f7} shows the ratio of the quartic to
quadratic fluctuations of the baryon number. We find that this ratio
approaches 1 at low temperature, which is consistent with the
lattice results as well as the calculations in the hadron resonance
gas model~\cite{Cheng2009,Karsch2003}. An interesting thing is that
we find a pronounced cusp in the ratio $\chi_{4}^{B}/\chi_{2}^{B}$
at the critical temperature in our calculations. However, whether
the cusp exists in the lattice results is not clear, because the
errors are very large at low temperature as the figure shows. We
have emphasized that answering the question whether there is a cusp
in the ratio of the quartic to quadratic fluctuations is very
important, because this ratio is a valuable probe of the
deconfinement and chiral phase transitions and have received lots of
attentions recent years~\cite{Karsch2007,Stokic2009}. Our
calculations indicate that the cusp in the ratio
$\chi_{4}^{B}/\chi_{2}^{B}$ is prominent even with physical strange
quark mass $m_{0}^{s}=140.7\;\mathrm{MeV}$ and relatively larger
(comparing with the physical value) light quark mass
$m_{0}^{l}=14.0\;\mathrm{MeV}$.

\begin{figure}[!htb]
\includegraphics[scale=1.2]{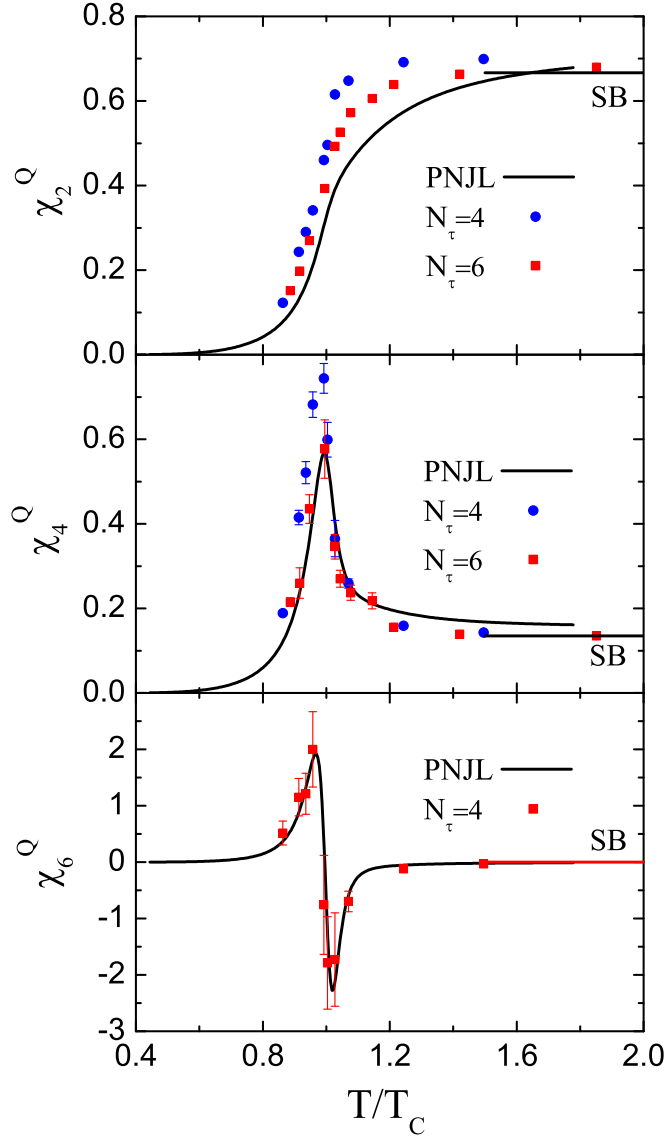}
\caption{(color online). Quadratic (top), quartic (middle) and the
6th order (bottom) fluctuations of electric charge as functions of
the temperature obtained in the PNJL model with
$m_{0}^{l}=14.0\;\mathrm{MeV}$ and $m_{0}^{s}=140.7\;\mathrm{MeV}$
and in the QCD simulations on $16^{3}\times4$ and $24^{3}\times6$
lattices in Ref.~\cite{Cheng2009}.}\label{f8}
\end{figure}

In Fig.~\ref{f8}, we show the quadratic, quartic and the 6th order
fluctuations of electric charge versus temperature obtained in our
calculations and lattice simulations. Once more, we find that the
electric charge quadratic fluctuations increase monotonically with
the temperature and the quartic fluctuations of the electric charge
present a prominent cusp at the critical temperature. As for the 6th
order fluctuations of electric charge, we find a pronounced
oscillation happening during the deconfinement and chiral phase
transitions. $\chi_{6}^{Q}$ is vanishing when the temperature is
below about $0.8T_{C}$, and with the temperature being increased
beyond $0.8T_{C}$, the 6th order fluctuations increase rapidly. When
the temperature is near $T_{C}$, $\chi_{6}^{Q}$ changes from a
positive value to a negative one abruptly. Then, $\chi_{6}^{Q}$
increases with the temperature and approaches the Stefan-Boltzmann
value ($\chi_{6}^{Q,SB}=0$) from below. From the bottom panel of
Fig.~\ref{f8}, One can find that the 6th order fluctuations of
electric charge are almost odd functions with respective to
$T-T_{C}$. Comparing Fig.~\ref{f8} with Fig.~\ref{f7}, we find that
the fluctuations of electric charge calculated in the PNJL model are
better consistent with those obtained in lattice calculations than
the fluctuations of the baryon number. This is because more
contributions to the electric charge fluctuations come from the
light quarks, since the absolute value of the electric charge of $u$
quarks 2/3 is larger than that of $s$ quarks $1/3$.

\begin{figure}[!htb]
\includegraphics[scale=0.5]{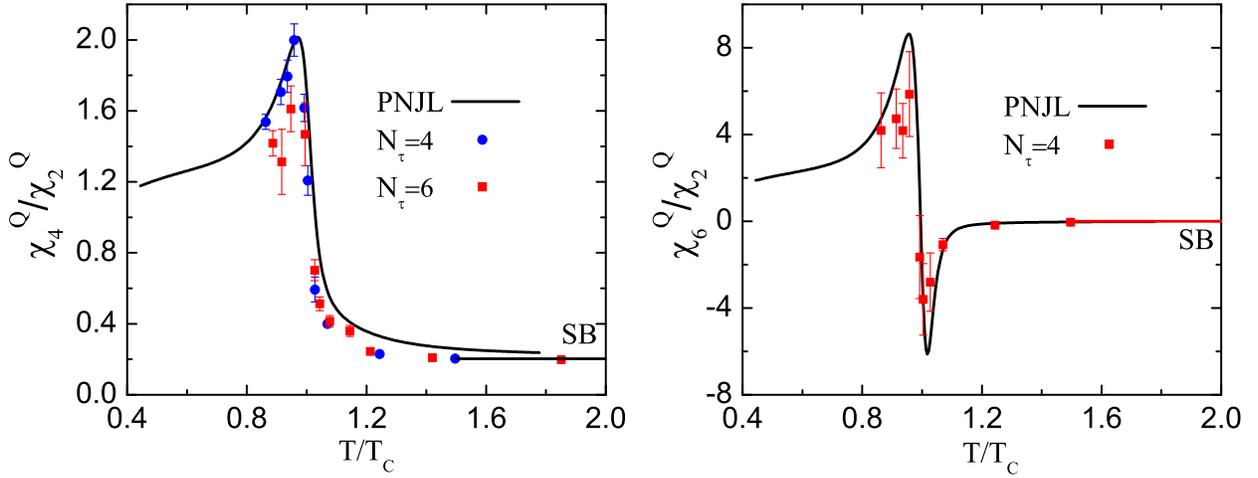}
\caption{(color online). Left panel: Ratio of the quartic to
quadratic electric charge fluctuations versus temperature obtained
in the PNJL model with $m_{0}^{l}=14.0\;\mathrm{MeV}$ and
$m_{0}^{s}=140.7\;\mathrm{MeV}$ and in the QCD simulations on
$16^{3}\times4$ and $24^{3}\times6$ lattices in
Ref~\cite{Cheng2009}. Right panel: Ratio of the 6th to 2nd order
fluctuations of electric charge versus temperature calculated in the
PNJL model with $m_{0}^{l}=14.0\;\mathrm{MeV}$ and
$m_{0}^{s}=140.7\;\mathrm{MeV}$ and in the QCD simulations on
$16^{3}\times4$ lattices in Ref.~\cite{Cheng2009}.}\label{f9}
\end{figure}

In Fig.~\ref{f9}, we show the ratios of the 4th to 2nd order and the
6th to 2nd order fluctuations of electric charge as functions of the
temperature obtained in the PNJL model and the lattice simulations.
One can find that the results given by the PNJL model are well
consistent with those obtained in the lattice simulations. Comparing
$\chi_{4}^{Q}/\chi_{2}^{Q}$ with $\chi_{4}^{B}/\chi_{2}^{B}$ in the
bottom panel of Fig.~\ref{f7}, we find there is a prominent cusp at
the critical temperature in the curve of $\chi_{4}^{Q}/\chi_{2}^{Q}$
as same as that in the curve of $\chi_{4}^{B}/\chi_{2}^{B}$.
However, different from $\chi_{4}^{B}/\chi_{2}^{B}$, at low
temperature $\chi_{4}^{Q}/\chi_{2}^{Q}$ is larger than unity and is
a raising function of the temperature, since it is expected that at
low temperature quartic fluctuations of electric charge are enhanced
relative to the quadratic fluctuations~\cite{Cheng2009}. As for the
ratio $\chi_{6}^{Q}/\chi_{2}^{Q}$, we find it oscillates violently
at the phase transition temperature.

\begin{figure}[!htb]
\includegraphics[scale=0.5]{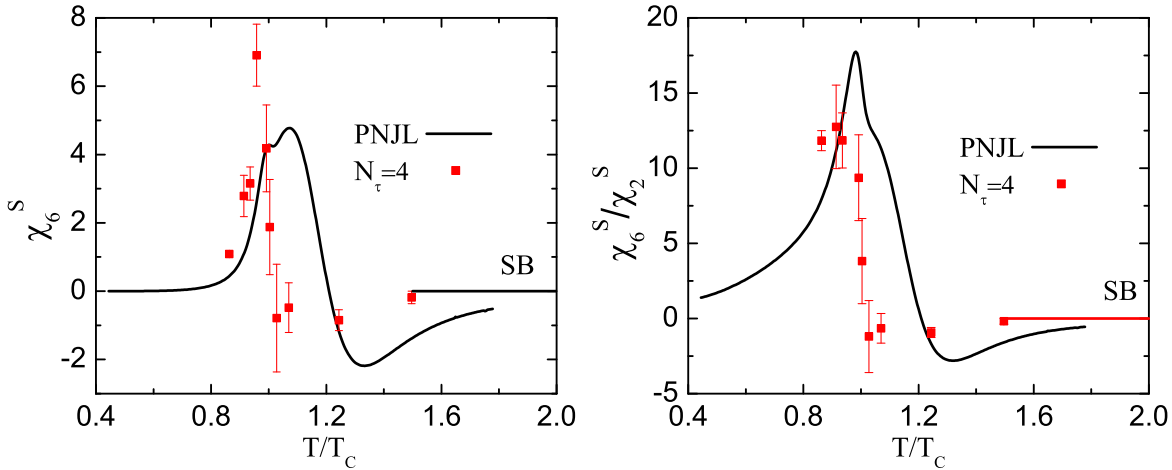}
\caption{(color online). The 6th order fluctuations of strangeness
(left panel) and the ratio $\chi_{6}^{S}/\chi_{2}^{S}$ (right panel)
as functions of the temperature calculated in the PNJL model with
$m_{0}^{l}=14.0\;\mathrm{MeV}$ and $m_{0}^{s}=140.7\;\mathrm{MeV}$
and in the QCD simulations on $16^{3}\times4$ lattices in
Ref.~\cite{Cheng2009}.}\label{f10}
\end{figure}

We have discussed the quadratic and quartic fluctuations of the
strange quarks in details above. It is easily verified that the
fluctuations of strange quarks are identical to those of
strangeness. Therefore, we will focus on the 6th order fluctuations
of strangeness in the following. Fig.~\ref{f10} shows the 6th order
fluctuations of strangeness and the ratio
$\chi_{6}^{S}/\chi_{2}^{S}$ as functions of the temperature
calculated in the PNJL model, which are also compared with the QCD
simulations on $16^{3}\times4$ lattices in Ref~\cite{Cheng2009}. We
find that $\chi_{6}^{S}$ and $\chi_{6}^{S}/\chi_{2}^{S}$ obtained in
the PNJL model are qualitatively consistent with those obtained in
lattice simulations , i.e. $\chi_{6}^{S}$ and
$\chi_{6}^{S}/\chi_{2}^{S}$ oscillate during the deconfinement and
chiral phase transitions. However, there are some quantitative
differences between these two approaches: the temperature where
$\chi_{6}^{S}$ and $\chi_{6}^{S}/\chi_{2}^{S}$ vanish during the
oscillation in the PNJL model is shifted from $T_{C}$ to about
$1.2T_{C}$ with respective to the results in the lattice
simulations. Similar phenomena have been found in the calculations
of $\chi_{4}^{s}$ and $\chi_{4}^{s}/\chi_{2}^{s}$ above in
Fig.~\ref{f6}, since much larger current mass of the strange quarks
results in that the chiral restoration phase transition for the
strange quark sector occurs at relative larger temperature than that
for the light quark sector as Fig.~\ref{f2} shows.

\begin{figure}[!htb]
\includegraphics[scale=1.2]{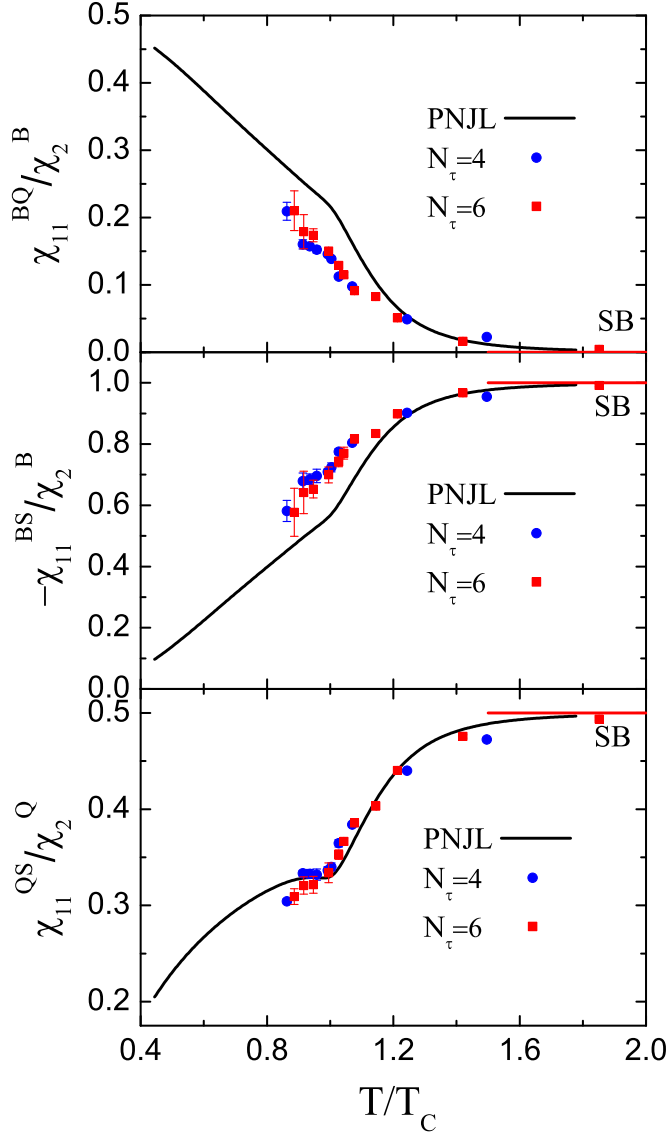}
\caption{(color online). Correlations of baryon number and electric
charge normalized to quadratic fluctuations of baryon number (top),
baryon number and strangeness to quadratic fluctuations of baryon
number (middle), and electric charge and strangeness to quadratic
fluctuations of electric charge (bottom) as functions of the
temperature obtained in the PNJL model with
$m_{0}^{l}=14.0\;\mathrm{MeV}$ and $m_{0}^{s}=140.7\;\mathrm{MeV}$
and in the QCD simulations on $16^{3}\times4$ and $24^{3}\times6$
lattices in Ref.~\cite{Cheng2009}.}\label{f11}
\end{figure}

In Fig.~\ref{f11}, we show correlations among conserved charges, in
more details, including the correlations of baryon number and
electric charge, baryon number and strangeness and those of electric
charge and strangeness. The former two correlations are normalized
to the quadratic fluctuations of baryon number and the last one is
to quadratic fluctuations of electric charge. One can find that
$\chi_{11}^{BQ}/\chi_{2}^{B}$ calculated in the PNJL model decreases
with the increase of the temperature, which agrees with the
calculations in lattice simulations. When the temperature is about
$1.5T_{C}$, this ratio approaches zero, the limit value of
noninteracting massless quark gas, since the sum of electric charges
of light quarks and strange quarks vanishes in this limit. Ones can
also find that our calculated results of
$\chi_{11}^{BQ}/\chi_{2}^{B}$ are a little larger than those of
lattice calculations at the critical temperature, but well
consistent with the lattice results at high temperature. The
correlation between baryon number and strangeness has been thought
as a useful diagnostic of strongly interacting matter, which can be
extracted not only theoretically from lattice QCD and effective
model calculations but also experimentally from event-by-event
fluctuations~\cite{Koch2005}. We find that the correlation of baryon
number and strangeness approaches zero at low temperature, since the
strange hadrons (here in the PNJL model are the three quark clusters
with strange quarks) are much heavier to excite in the
thermodynamical system. The ratio $-\chi_{11}^{BS}/\chi_{2}^{B}$
calculated in the PNJL model increases with the temperature and
approaches the Stefan-Boltzmann value at about $1.5T_{C}$, which
agrees with the lattice QCD simulations. Same as the correlation of
baryon number and strangeness, the correlation of strangeness and
electric charge is also suppressed at low temperature and the ratio
$\chi_{11}^{QS}/\chi_{2}^{Q}$ increase with the temperature. One can
find that our calculated results of $\chi_{11}^{QS}/\chi_{2}^{Q}$
are well consistent with those of the lattice simulations. An
interesting thing is that both the results given by the two
approaches indicate that there is a plateau in the ratio of
$\chi_{11}^{QS}/\chi_{2}^{Q}$ at the critical temperature.

\section{Summary and Discussions}

In this work, we have studied the fluctuations and correlations of
the conserved charges, i.e. the baryon number, electric charge and
the strangeness, in the Polyakov--Nambu-Jona-Lasinio (PNJL) model at
finite temperature. We have also compared our calculated results
with those obtained from the recent lattice calculations performed
with an improved staggered fermion action at two values of the
lattice cutoff with almost physical up and down quark masses and a
physical value for the strange quark mass~\cite{Cheng2009}.

In order to study the fluctuations of conserved charges much better,
we have calculated the fluctuations of light quarks (up and down
quarks) and strange quarks before the calculations for the conserved
charges. Our calculations indicate that the quadratic fluctuations
of light quarks, strange quarks and conserved charges (including the
baryon number, electric charge and the strangeness) increase from
zero at low temperature to their corresponding Stefan-Boltzmann
values at high temperature, since particles excited at finite
temperature are increased with the increase of temperature; the
quartic fluctuations are characterized by an pronounced cusp (the
cusp is high and sharp for light quarks and low and broad for
strange quarks, but both are prominent) during the deconfinement and
chiral phase transitions; the 6th order fluctuations for the quarks
and conserved charges oscillate with the occurrence of the phase
transitions. These qualitative features shown in our calculations
are well consistent with those given by the lattice calculations,
which confirms that the effective model (here is the PNJL model)
captures the right symmetry of the QCD, i.e. the chiral symmetry
group, since different phase transitions are govern by different
symmetry groups.

Comparing our calculated results with those obtained in lattice QCD
simulations quantitatively, we have found that the fluctuations of
light quarks obtained in the PNJL model are well consistent with
those calculated in the lattice simulations. However, for the
fluctuations of strange quarks, there are discrepancies between
these two approaches. In more details, calculations of the PNJL
model indicate that the cusp of the quartic fluctuations and the
oscillation of the 6th order fluctuations for strange quarks are
located at about $1.2T_{C}$, while they are still located at $T_{C}$
as same as the light quarks in the lattice calculations in
Ref.~\cite{Cheng2009}. The reason for this discrepancy is that the
pseudo-critical temperature of the chiral restoration phase
transition for strange quarks is a little larger than that for light
quarks in the PNJL model due to the larger current quark mass of
strange quarks~\cite{Fu2008}, and locations of the cusp of the
quartic fluctuations and the oscillation of the 6th order
fluctuations for strange quarks is related with this pseudo-critical
temperature for strange quarks, which is about $1.2T_{C}$ in the
PNJL model. On the contrary, results obtained in lattice
calculations in Ref.~\cite{Cheng2009} indicate that the
pseudo-critical temperature of the strange quark chiral phase
transition is same as that for light quarks. However, we should
emphasize that in some other lattice simulations, for example in
Ref.~\cite{Aoki2006b}, these two pseudo-critical temperatures are
different and the pseudo-critical temperature for strange quarks is
larger than that for light quarks. Furthermore, because of the
effects of current quark masses in the PNJL model, especially for
the strange quarks, fluctuations obtained in the PNJL model
approaches to their corresponding Stefan-Boltzmann values more
slowly than those obtained in lattice calculations. Comparing the
fluctuations of electric charge with those of baryon number, we find
that the former calculated in the PNJL model are better consistent
with those obtained in lattice calculations than the latter, since
the fluctuations of electric charge have more contributions from
light quarks than the fluctuations of baryon number. Although the
conclusion about whether there is a cusp during the phase transition
in the ratio of quartic to quadratic fluctuations of the baryon
number, i.e. $\chi_{4}^{B}/\chi_{2}^{B}$, is not clear in lattice
calculations due to the large errors in their simulations, We indeed
find a pronounced cusp in this ratio $\chi_{4}^{B}/\chi_{2}^{B}$ at
the critical temperature in our calculations, which confirms that
$\chi_{4}^{B}/\chi_{2}^{B}$ is a valuable probe of the deconfinement
and chiral phase transitions.

Furthermore, we have also studied the correlations among conserved
charges in the PNJL model and compared our calculated results with
those obtained in lattice simulations. Same as the fluctuations of
conserved charges, except for some small quantitative differences,
these correlations among conserved charges calculated in the PNJL
model are well consistent with those obtained in the lattice
calculations.

\section*{Acknowledgements}

This work was supported by the National Natural Science Foundation
of China under Contract Nos. 10425521, 10675007, and 10935001, and
the Major State Basic Research Development Program under Contract
No. G2007CB815000. One of the authors (W. J. F.) also acknowledges
financial support from China Postdoctoral Science Foundation No.
20090460534.


\begin{thebibliography}{}


\bibitem{Shuryak2004}
E.~V.~Shuryak, Prog. Part. Nucl. Phys. {\bf 53}, 273 (2004).

\bibitem{Gyulassy2005}
M.~Gyulassy, and L.~McLerran, Nucl. Phys. {\bf A 750}, 30 (2005).

\bibitem{Shuryak2005}
E.~V.~Shuryak, Nucl. Phys. {\bf A 750}, 64 (2005).

\bibitem{Arsene2005}
I.~Arsene \textit{et al}, Nucl. Phys. {\bf A 757}, 1 (2005).

\bibitem{Back2005}
B.~B.~Back \textit{et al}, Nucl. Phys. {\bf A 757}, 28 (2005).

\bibitem{Adams2005}
J.~Adams \textit{et al}, Nucl. Phys. {\bf A 757}, 102 (2005).

\bibitem{Adcox2005}
K.~Adcox \textit{et al}, Nucl. Phys. {\bf A 757}, 184 (2005).

\bibitem{Blaizot2007}
J.-P.~Blaizot, J. Phys. {\bf G 34}, S243 (2007).

\bibitem{Weber2005} F. Weber,
Prog. Part. Nucl. Phys. {\bf 54}, 193 (2005).

\bibitem{Alford2007} M. Alford, D. Blaschke, A. Drago, T. Kl\"{a}hn,
G. Pagliara, J. Shaffner-Bielich, Nature {\bf 445}, E 7 (2007).

\bibitem{Alford2008}
M.~Alford, A.~Schmitt, K.~Rajagopal, and T. Sch\"{a}fer, Rev. Mod.
Phys. {\bf 80}, 1455 (2008).

\bibitem{Fu2008b}
W.~J.~Fu, H.~Q.~Wei, and Y.~X.~Liu, Phys. Rev. Lett. {\bf 101},
181102 (2008).

\bibitem{Boyd1996}
G.~Boyd, J.~Engels, F.~Karsch, E.~Laermann, C.~Legeland,
M.~L\"{u}gemeier, and B.~Petersson, Nucl. Phys. {\bf B 469}, 419
(1996); J.~Engels, O.~Kaczmarek, F.~Karsch, and E.~Laermann, Nucl.
Phys. {\bf B 558}, 307 (1999).

\bibitem{Fodor2002}
Z.~Fodor, and S.~D. Katz, Phys. Lett. {\bf B 534}, 87 (2002); {\it
ibid}, J. High Energy Phys. {\bf 0203}, 014 (2002); Z.~Fodor, S.~D.
Katz, and K.~K. Szabo, Phys. Lett. {\bf B 568}, 73 (2003).

\bibitem{Allton2002}
C.~R. Allton, S. Ejiri, S. J. Hands, O. Kaczmarek, F. Karsch, E.
Laermann, Ch. Schmidt, and L. Scorzato, Phys. Rev. {\bf D 66},
074507 (2002); C.~R. Allton, S. Ejiri, S. J. Hands, O. Kaczmarek, F.
Karsch, E. Laermann, and Ch. Schmidt, Phys. Rev. {\bf D 68}, 014507
(2003); C.~R. Allton, M. D\"{o}ring, S. Ejiri, S. J. Hands, O.
Kaczmarek, F. Karsch, E. Laermann, and K. Redlich, Phys. Rev. {\bf D
71}, 054508 (2005).

\bibitem{Laermann2003}
E.~Laermann, and O.~Philipsen,
Ann. Rev. Nucl. Part. Sci. {\bf 53}, 163 (2003).

\bibitem{deForcrand2003}
P.~de Forcrand, and O.~Philipsen,
Nucl. Phys. {\bf B 642}, 290 (2002); {\bf B 673}, 170 (2003); P. de
Focrand, and S. Kratochvila, Nucl. Phys. B, Proc. Suppl. {\bf 153},
62 (2006).

\bibitem{Kratochvila0456}
S. Kratochvila, and P. de Focrand, Nucl. Phys. B, Proc. Suppl. {\bf
129}, 533 (2004); {\bf 140}, 514 (2005); Phys. Rev. {\bf D 73},
114512 (2006).

\bibitem{Gavai2005}
R.~V.~Gavai, and S.~Gupta, Phys. Rev. {\bf D 72}, 054006 (2005).

\bibitem{Gavai2006}
R.~V.~Gavai, and S.~Gupta, Phys. Rev. {\bf D 73}, 014004 (2006).


\bibitem{Aoki2006a}
Y.~Aoki, G.~Endrodi, Z.~Fodor, S.~D.~Katz, and K.~K.~Szab\'{o},
Nature {\bf 443}, 675 (2006).

\bibitem{Aoki2006b}
Y.~Aoki, Z.~Fodor, S.~D.~Katz, and K.~K.~Szab\'{o}, Phys. Lett. {\bf
B 643}, 46 (2006).

\bibitem{Cheng2006}
M.~Cheng \textit{et al.}, Phys. Rev. {\bf D 74}, 054507 (2006).


\bibitem{Ejiri2006a}
S.~Ejiri, F.~Karsch, and K.~Redlich, Phys. Lett. {\bf B 633}, 275
(2006).

\bibitem{Ejiri2006b}
S.~Ejiri \textit{et al.}, Nucl. Phys. {\bf A 774}, 837 (2006).

\bibitem{Karsch2006}
F.~Karsch, S.~Ejiri, and K.~Redlich, Nucl. Phys. {\bf A 774}, 619
(2006).

\bibitem{Karsch2007}
F.~Karsch, PoS CPOD07, 026 (2007).

\bibitem{Cheng2009}
M.~Cheng \textit{et al.}, Phys. Rev. {\bf D 79}, 074505 (2009).

\bibitem{Ejiri2009}
S.~Ejiri \textit{et al.}, arXiv:0909.5122 [hep-lat].

\bibitem{Stephanov1998}
M.~A.~Stephanov, K.~Rajagopal, and E.~V.~Shuryak, Phys. Rev. Lett.
{\bf 81}, 4816 (1998).

\bibitem{Hatta2003}
Y.~Hatta, and T.~Ikeda, Phys. Rev. {\bf D 67}, 014028 (2003).


\bibitem{Jeon2004}
S.~Jeon, and V.~Koch, in: R.~C.~Hwa, X.~N.~Wang (Eds.), Quark Gluon
Plasma, vol. 3, World Scientific Publishing, 2004, p.430.

\bibitem{Stephanov2004}
M.~Stephanov, Acta Phys. Pol. B {\bf 35}, 2939 (2004).

\bibitem{Jeon2000}
S.~Jeon, and V.~Koch, Phys. Rev. Lett. {\bf 85}, 2076 (2000).

\bibitem{Koch2005}
V.~Koch, A.~Majumder, and J.~Randrup, Phys. Rev. Lett. {\bf 95},
182301 (2005).

\bibitem{Stokic2009}
B.~Stoki\'{c}, B.~Friman, and K.~Redlich, Phys. Lett. {\bf B 673},
192 (2009).

\bibitem{Abelev2009}
B.~I.~Abelev \textit{et al.} (STAR Collaboration), Phys. Rev. Lett.
{\bf 103}, 092301 (2009).


\bibitem{Fu2008}
W.~J.~Fu, Z.~Zhang, and Y.~X.~Liu, Phys. Rev. {\bf D 77}, 014006
(2008).


\bibitem{Meisinger9602}
P.~N.~Meisinger, and M.~C.~Ogilvie, Phys. Lett. {\bf B 379}, 163
(1996); P.~N.~Meisinger, T.~R.~Miller, and M.~C.~Ogilvie, Phys. Rev.
{\bf D 65}, 034009 (2002).

\bibitem{Pisarski2000}
R.~D.~Pisarski, Phys. Rev. {\bf D 62}, 111501 (2000); A.~Dumitru and
R.~D.~Pisarski, Phys. Lett. {\bf B 504}, 282 (2001); Phys. Lett.
{\bf B 525}, 95 (2002); Phys. Rev. {\bf D 66}, 096003 (2002).

\bibitem{Fukushima2004}
K.~Fukushima, Phys. Lett. {\bf B 591}, 277 (2004).


\bibitem{Ratti2006a}
C.~Ratti, M.~A.~Thaler, and W.~Weise, Phys. Rev. {\bf D 73}, 014019
(2006).

\bibitem{Ratti2006b}
S.~R\"{o}{\ss}ner, C.~Ratti, and W.~Weise, Phys. Rev. {\bf D 75},
034007 (2007).


\bibitem{Ciminale2008}
M.~Ciminale, R.~Gatto, N.~D.~Ippolito, G.~Nardulli, and M.~Ruggieri,
Phys. Rev. {\bf D 77}, 054023 (2008).

\bibitem{Zhang2006}
Z.~Zhang, and Y.~X.~Liu, Phys. Rev. {\bf C 75}, 064910 (2007)
(arXiv: hep-ph/0610221).

\bibitem{Fu2009}
W.~J.~Fu, and Y.~X.~Liu, Phys. Rev. {\bf D 79}, 074011 (2009).

\bibitem{Ghosh2006}
S.~K.~Ghosh, T.~K.~Mukherjee, M.~G.~Mustafa, and R.~Ray, Phys. Rev.
{\bf D 73}, 114007 (2006); S.~Mukherjee, M.~G.~Mustafa, and R.~Ray,
Phys. Rev. {\bf D 75}, 094015 (2007)


\bibitem{Rehberg1996}
P.~Rehberg, S.~P.~Klevansky, and J.~H\"{u}fner, Phys. Rev. {\bf C
53}, 410 (1996).


\bibitem{Amsler2008}
C.~Amsler \textit{et al.} (Particle Data Group),  Phys. Lett. {\bf B
667}, 1 (2008).

\bibitem{Karsch2003}
F.~Karsch, K.~Redlich, and A.~Tawfik, Phys. Lett. {\bf B 571}, 67
(2003).

\end{thebibliography}
\end{document}